\documentstyle[12pt]{article}
\input epsf

\begin{document}
\begin{titlepage}
\thispagestyle{empty}
\title{
\vspace*{-1cm} 
\begin{flushright}
{\small KEK-TH-755}\\
\end{flushright}
\vspace{2.0cm}
Study of $\gamma\pi \rightarrow \pi\pi$ below 1 GeV using Integral Equation Approach}
\vspace{4.0cm}
\author{Tran N. Truong \\
\small \em Centre de Physique Th{\'e}orique, 
{\footnote {unit{\'e} propre 014 du
CNRS}}\\ 
\small \em Ecole Polytechnique \\
\small \em F91128 Palaiseau, France \\
 and \\ 
\small \em Institute of  Particle and Nuclear Studies, \\
\small \em High Energy Accelerator Research Organization,\\
\small \em 1-1 Oho, Tsukuba, Ibaraki, 305 - 0801 Japan }

\date{October 2001}

\maketitle

\begin{abstract}
The scattering of $\gamma  \pi \to \pi \pi$ is studied using the axial anomaly, 
elastic unitarity, analyticity and crossing symmetry. Using the technique to derive the Roy's equation,  an integral  equation
for the P-wave amplitude is obtained in terms of the strong P-wave pion pion phase shifts.
Its solution is obtained numerically by an iteration procedure using the starting point as the
solution of the integral equation of the Muskelshsvilli-Omnes type. It is, however, ambiguous and 
depends  sensitively on the second derivative of the P-wave amplitude  at
$s=m_\pi^2$ which cannot  directly be  measured.

\end{abstract}

\vspace{2.0cm}


\end{titlepage}

\section{Introduction}
\vskip 0.25 in

One of the fundamental calculation in particle theory is the $\pi^0 \to \gamma
\gamma$ decay rate \cite{adler1}. It is a combination of Partial
 Conserved Axial Current (PCAC) and 
the short distance behavior  of Quantum Chromodynamics (QCD):
\begin{equation}
A(\pi^0 \rightarrow \gamma\gamma)=iF_{\gamma\gamma}\epsilon^{\mu\nu\sigma\tau}
\epsilon_{\mu}^{\ast} k_{\nu}\epsilon_{\sigma}^{\ast}k_{\tau}^{\prime}\label{eq:pgg}
\end{equation}
with
\begin{equation}
F_{\gamma\gamma}=\frac{e^2 N_c}{12\pi^2f_\pi}=0.025 GeV^{-1} \label{eq:npgg}
\end{equation}
where $e$ is the electric charge, $f_\pi= 0.0924$ GeV and
 $N_c=3$ is the number of color in QCD.
This calculation is valid in a world where 
the $\pi^0$ is massless. Some corrections have to be made in order
 to take into account of the finite value of the pion mass. It turns out that the
 massless pion anomaly formula is in a very good agreement with the pion life time data \cite{pdg},
implying that  the correction due to the physical pion
 mass in 
Eq. (\ref{eq:npgg}) is very small.

Another Axial Anomaly result is the process $\gamma \pi \to \pi\pi$ or
 its analytical continuation $\gamma \to 3\pi$ \cite{adler2}.
 This last process requires  more
 corrections because, for practical consideration, measurements are done
at an energy  
far from the chiral limit where the anomaly formula is applicable. Furthermore 
the analytical continuation from one process to the other is a delicate procedure due to 
the presence of the complex singularity which is absent in the former reaction.
The calculation of the process $\gamma \pi \to \pi\pi$ is
 in itself interesting because of future experiments  being
proposed at various accelerator facilities and also of
 its important role in the calculation of $\pi^0 \to \gamma \gamma^*$ \cite{truong1}.

The  $\gamma \pi \to \pi\pi$ amplitude is given as:
\begin{equation}
A(\gamma(k)\pi^0(p_0)\rightarrow
\pi^+(p_1)\pi^-(p_2))=i\epsilon^{\mu\nu\sigma\tau}\epsilon_\mu
 p_{0\nu} p_{1\sigma}p_{2\tau}G(s,t,u) \label{eq:anomaly2}
\end{equation}
where s, t, u are the kinematical variables for this process and will be defined below.

In the  chiral limit (the zero limit of the pion 4-momenta), the matrix 
element is given by the anomaly equation:
\begin{equation}
G(0,0,0) \equiv \lambda = \frac{e}{4\pi^2 f_\pi^3}=  9.70  GeV^{-3} \label{eq:anomaly3}
\end{equation}
where the zero in the argument of $G(0,0,0)$ refers to the chiral
 limit of the massless pions; the
 number of colors $N_{c}$ is equal to 3.

Experimentally, $\lambda$ is measured at an average photon pion 
energy of 0.4 GeV  and assuming that there is no momentum dependence
in G(s, t, u), it is equal to
\cite{antipov}:
\begin{equation}
\lambda^{expt}=12.9\pm0.9\pm0.5  GeV^{-3} \label{eq:anomaly4}
\end{equation}

The agreement between experiment and theory is not  good but certainly corrections will have to be
made because  measurements made in this experiment are far from the chiral symmetry limit. 

The calculations of this process are usually done within the Vector Meson Dominance models (VMD)
\cite{rudaz, bando,  sharp}.Recently this process is discussed within the framework of Chiral Perturbation
Theory (ChPT) to one loop \cite{bijnens} and also a combination of ChPT and VMD \cite{holstein} and
the unitarization of the ChPT two loop amplitude \cite{hannah}. 

The purpose of this paper is to investigate the scattering of $\gamma \pi^0 \to \pi^+ \pi^-$ using
dispersion relation, elastic unitarity and the knowledge of the P-wave
pion pion phase shits. An  integral equation is obtained and is
similar to the Muskelishvilli-Omnes (M-O) integral equation used in the pion
form factor  calculation \cite{omnes}. The integral equation obtained here is, however, much more complicated due
to the symmetry of the problem. Its solution can only be obtained by a numerical method. 

The  pion form factor calculation, using the M-O integral equation approach, yields a   
pion  radius too low by 10\% and a  modulus of the pion form factor at
the $\rho$ resonance also  too
low by 15\%. This is due to the assumption of the elastic unitarity relation which is only valid in the low
energy region below 1.2 GeV but cannot be true at and above the $\rho^\prime$( 1.5 GeV) region. In order to remove this discrepancy one
has to use also, as input, the pion r.m.s radius and hence, one  has either to make an extra subtraction in the dispersion relation or to  make  use of the polynomial ambiguity of the solution of
the M-O equation.  One would then get not only the correct value of the absolute value of the pion form
factor at the
$\rho$ mass (i.e. the
$\rho$ leptonic width), but also a complete agreement with the pion form factor below 1 GeV
\cite{truong1}. The phases of the form factor are of course the
experimental P-wave $\pi\pi$ phase shifts due to the solution of the
M-O equation. 

We also face the same problem in the calculation of the scattering
$\gamma \pi^0 \to \pi^+ \pi^-$. The
problem  could, however, be  more serious here than in the pion form factor
calculation due to the existence of the t and u channels . Not only the first derivative
of the  dominant P-wave amplitude vanishes at the energy squared
$s=m_\pi^2$, but its second derivative at this energy 
is not accessible to experiments because of the lack of experimental precision. 

A further complication is due to the ambiguity of the solution of the integral  equation
obtained here due to the symmetry of the problem. It is related to but
is not purely  the polynomial type. For this reason we cannot make a comparable  prediction of the $\gamma \pi^0 \to \pi^+ \pi^-$ cross section at the $\rho$ mass or
$\Gamma(\rho \to \pi\gamma)$ width. The measurement of  this width could then be used to make a prediction
of the energy dependence of the $\gamma \pi^0 \to \pi^+ \pi^-$ away from the $\rho$ mass and in
particular in the low energy  region where the the first measurement of $\lambda$, Eq.
(\ref{eq:anomaly4}), was made.

\section{ Kinematics and Partial Wave Projection}

The kinematics of this process are defined as: $s=(k+p_0)^2$, $t=(p_1-p_0)^2$ and $u=(p_2-p_0)^2$.
Because all particles involved are on shell, one has $s+t+u=3m_\pi^2$. In the center of mass
system, in terms of the scattering angle $\theta$, we have:
\begin{eqnarray}
t&=&\frac{3m_\pi^2-s}{2}+\frac{1}{2}(s-m_\pi^2)\sqrt{1-4m_\pi^2/s}\cos\theta \nonumber \\
u&=&\frac{3m_\pi^2-s}{2}-\frac{1}{2}(s-m_\pi^2)\sqrt{1-4m_\pi^2/s}\cos\theta \label{eq:tu} 
\end{eqnarray}
 The partial wave expansion
for
$G(s,t,u)$ is given as follows \cite{ hannah, gourdin}:
\begin{equation}
G(s,t,u) = \sum_{odd  l} G_l(s)P_l^{\prime}(\cos\theta) \label{eq:pw1}
\end{equation}
where $\theta$ is the scattering angle and $P_l^{\prime}$ is the first derivative of the Legendre
polynomial. Hence the lowest partial wave is:
\begin{equation}
G_1(s) = \frac{3}{8\pi}\int d\Omega \sin^2\theta G(s,t,u) \label{eq:pw2}
\end{equation}
In terms of the function $G_{3\pi}(s,t,u)$ the differential cross section for the process
  $\gamma \pi^0
\rightarrow \pi^+\pi^-$ is 
\begin{equation}
\frac{d\sigma}{d\cos\theta} =
\frac{1}{1024\pi}(s-m_\pi^2)\frac{(s-4m_\pi^2)^{3/2}}{s^{1/2}}\sin^2\theta\mid G(s,t,u)\mid^2
\label{eq:x}
\end{equation}

\section{Vector Meson Dominance and Pion Form Factor}

Because our integral equation solution is a more sophisticated and
precised approach to the vector meson dominance model (VMD)
\cite{sakurai, wagner}, where unitarity and dispersion relation are extensively used, it cannot avoid
the same problems which are presented in these models. Namely the solution can only uniquely obtained
when the asymptotic behavior of the solution is specified.  It is then
useful to  review briefly the
VMD models and calculations of the pion form factor with and without introducing the contact term.

\subsection { Vector Meson Dominance Models for $\gamma
\pi^0 \to \pi^+\pi^-$ process }
Let us consider the VMD models without and with the contact terms for the $\gamma
\pi^0 \to \pi^+\pi^-$ process  as previouly discussed in the
litterature
\cite{rudaz, bando}. Without the contact term, the VMD model for 
$\gamma\pi^0
\rightarrow
\pi^+\pi^-$ amplitude is:
\begin{equation}
G^{vmd}(s,t,u) =\frac{\lambda}{3}(\frac{m_\rho^2}{m_\rho^2-s}+\frac{m_\rho^2}{m_\rho^2-t}+
\frac{m_\rho^2}{m_\rho^2-u}) \label{eq:3p1}
\end{equation}
where $s,t,u$ are the  invariant kinematics and $\lambda$ is defined by Eq. (\ref{eq:anomaly3}). With a contact term, it  can be written as:
\begin{equation}
G^{vmdc}(s,t,u)
=\frac{\lambda}{3-c}\{\frac{m_\rho^2}{m_\rho^2-s}+\frac{m_\rho^2}{m_\rho^2-t}+
\frac{m_\rho^2}{m_\rho^2-u}-c\} \label{eq:3p2}
\end{equation}
where $c$ is proportional to the strength of the contact term.
Eq. (\ref{eq:3p2}) can be rearranged to give:
\begin{equation}
G^{vmdcc}(s,t,u)
=\frac{\lambda}{3}\{[\frac{m_\rho^2}{m_\rho^2-s}(1+\frac{c}{3-c}\frac{s}{m_\rho^2})]+[s\to t] +[s
\to u]\}
\label{eq:3p3}
\end{equation}
In Eq. (\ref{eq:3p1}), the $\gamma\pi^0\rightarrow \pi^+\pi^-$ amplitude vanishes as $s,t,u \to
\infty$ while those in Eqs. (\ref{eq:3p2}, \ref{eq:3p3}) do not vanish due to the presence of the
contact term. We have introduced phenomenologically the contact term c
in the scattering $\gamma\pi^0\rightarrow \pi^+\pi^-$ without worrying
how it influences the VMD for the corresponding process
$P\to \gamma\gamma$. Assuming  a complete VMD for $P \to \gamma\gamma$, one has c=1 in order
that the KSRF relation \cite{KSRF} remains valid \cite{rudaz, bando}.

The strength of the contact term also influences the value of the
second derivative of the P-wave amplitude at $s=m_\pi^2$. Expand 
Eq.(\ref{eq:3p2}) in a
power series of  s,t,u:
\begin{equation}
 G^{vmdc}(s,t,u) = \lambda [1+\frac{3}{3-c}\frac{m_\pi^2}{m_\rho^2}+
\frac{1}{3-c}\frac{s^2+t^2+u^2}{m_\rho^4}+...] \label{eq:expand1}
\end{equation}
and the P-wave projection of this equation is:
\begin{equation}
 G_{1}(s) = \lambda [1+\frac{3}{3-c}\frac{m_\pi^2}{m_\rho^2}+
\frac{6}{5(3-c)}\frac{(s-m_\pi^2)^2}{m_\rho^4}+... ]\label{eq:expand2}
\end{equation}
where  the pion mass is introduced by hand. One has finally:
\begin{equation}
\frac{d^2 G_{1}(s)}{ds^2}\mid_{s=m_\pi^2}= \frac{12}{5(3-c)}\frac{1}{m_\rho^4} \lambda \label{eq:der2}
\end{equation}

Instead of characterizing the contact term by the infinite energy  behavior of the matrix element, we
can specify its presence by evaluating its second derivative for the P-wave at $s=m_\pi^2$. For the
pure VMD, $c=0$, it is equal to $(12/15)\lambda (m_\rho^{-4})$ and for the hidden
symmetry model \cite{bando},
$c=1$, it is
$(6/5)\lambda (m_\rho^{-4}) $.

  Eq. (\ref{eq:3p1}) yields a decay width
$\Gamma(\rho
\to
\pi\gamma)=36 KeV$ which is too small compared with the experimental value. Eq. (\ref{eq:3p2}) for
$c=1$ yields a decay width
$\Gamma(\rho
\to
\pi\gamma)=81 KeV$ in much better agreement with the data (see below). The
experimental value of the $\rho \pi\pi$ coupling or the KSRF
relation \cite{KSRF} for the $\rho\pi\pi$ coupling are used to calculate these widths.

One can improve these equations by making the vector meson $\rho$ unstable using the self-energy
correction for the $\rho$ propagator \cite{gounaris} and  using the KSRF relation \cite{KSRF}. This same result can also be obtained using the inverse amplitude for the
vector form factor {\em without} assuming the KSRF relation. The $\rho$ width obeys the KSRF relation as a
consequence of the implementation of the  unitarity relation \cite{truong2}.  The factor
$m_\rho^2/(m_\rho^2-s)$ is then replaced by a function $\Omega(s)$
which is normalized to unity at s=0 and is defined as follows:

\begin{equation}
         \Omega(s) = \frac{1} {1 -s/s_{R} - {1\over
96\pi^2f_\pi^2}\{(s-4m_\pi^2)
 H_{\pi\pi}({s}) + {2s/3}\}} \label{eq:vu1}
\end{equation}
where $f_\pi=0.093 GeV$, and $s_{R}$ is related to the $\rho$ mass squared $m_\rho^2=0.593 GeV^2$ by 
requiring that the real part of the denominator of Eq. (\ref{eq:vu1}) vanishes at the $\rho$
mass; $H_{\pi\pi}({s})$ is a well-known integral over the phase space factor:
\begin{eqnarray}
H_{\pi\pi}(s) & = & 2-2\sqrt{\frac{s-4m_\pi^2}{s}}\ln\frac{\sqrt{s}+\sqrt{s-4m_\pi^2}}{2
m_\pi} +i\pi\sqrt{\frac{s-4m_\pi^2}{s}} ;  s\geq 4m_\pi^2 \nonumber \\
   & = & 2-2\sqrt{\frac{4m_\pi^2-s}{s}}\arctan{\sqrt{\frac{s}{s-4m_\pi^2}}} ; 0\leq
 s\leq 4 m_\pi^2  \nonumber \\
     & = & 2-2\sqrt{\frac{s-4m_\pi^2}{s}}\ln{\frac{\sqrt{4m_\pi^2-s}+\sqrt{-s}}{2m_\pi}};
 s\leq 0 \label{eq:h}
\end{eqnarray}
Let us call the phase of $\Omega(s)$ $\delta$. Then $\Omega(s)$ has the following phase
representation:
\begin{equation}
\Omega(s)=\exp\frac{s}{\pi}\int_{4m_\pi^2}^\infty \frac{\delta(z)dz}{z(z-s-i\epsilon)}
\label{eq:o}
\end{equation}
The phase $\delta$ is exactly the elastic P-wave $\pi\pi$  phase shifts as can be seen from Fig.
1. Alternatively one can use the experimental phase shift to calculate  the function $\Omega(s)$ but
the expression given above is most convenient.

 Other functions
$\Omega(s)$ normalised to unity at
$s=s_0$ can be expressed in terms of
 the function $\Omega(s)$ by the simple relation $\Omega(s,s_0)=\Omega(s)/\Omega(s_0)$. 
 
 $\Omega(s)$ as given by Eq. (\ref{eq:vu1}) has a ghost pole at
 $s=-2.5.10^5 GeV^2$ which is far away
from the physical region relevant to our calculation
  and hence is irrelevant for our low energy calculation.

The function $\Omega(s)$ defined here is the same as the inverse of
the D-function given by reference\cite{holstein} except for the
definition of the $\rho$ mass which is approximate there.

In both approaches, the chiral symmetry limit should be defined as the
limit of s, t, u tend to zero first and then $m_\pi^2 \to 0$. This
order should be respected because the branch point at $s,t,u
=4m_\pi^2$ also goes to zero in the chiral limit. Using this
definition we could have calculated $\overline{\lambda}$ in terms of
$\lambda$ without using the large $N_c$ limit, but the difference is
negligible as discussed previously.

 Replacing $m_\rho^2/(m_\rho^2-s)$ by $\Omega(s)$ in Eq. (\ref{eq:3p1})
 yields 
$\Gamma(\rho\to\pi\gamma)=42 KeV$ and  with c=1 (the hidden symmetry model with an additional assumption of a
complete vector meson model for $\pi^0 \to \gamma \gamma$) in Eq. (\ref{eq:3p2})
  gives
$\Gamma(\rho\to\pi\gamma)=95 KeV$. The difference between these values
 and those obtained previouly are just due to the $\rho$ finite width
 correction. 
 These results show the importance  of the presence of the
contact term. While the present experimental data on the $\Gamma(\rho \to \pi\gamma)$ are not
settled, it is likeky that the result for the hidden symmetry model is favored (see below).

With chiral symmetry broken, the pions acquired a finite but small mass, Eqs. (\ref{eq:3p1},
\ref{eq:3p2},\ref{eq:3p3}) become, respectively:
\begin{equation}
G^{vmd}(s,t,u) =\frac{\lambda}{3}\{ \Omega(s) + \Omega(t) + \Omega(u) \} \label{eq:3p11}
\end{equation}
\begin{equation}
G^{vmdc}(s,t,u)2
=\frac{\lambda}{3-c}\{ \Omega(s) + \Omega(t) + \Omega(u) - c \} \label{eq:3p22}
\end{equation}
and 
\begin{equation}
G^{vmdcc}(s,t,u)
=\frac{\lambda}{3}\{ [\Omega(s)(1+\frac{c}{3-c}\frac{s}{m_\rho^2})]+[s\to t] +[s
\to u]\}
\label{eq:3p33}
\end{equation}

  Eqs. (\ref{eq:3p11},\ref{eq:3p22},\ref{eq:3p33}),   do not satisfy, 
however,  the elastic unitarity relation, i.e. the  projected P-wave   do not have the phase
of the P-wave
$\pi \pi$ interactions below 1 GeV \cite{watson} as can be shown in
Fig. 3. This result is not surprising because the multiple $\pi\pi$ scattering correction, which should be relevant for this
problem, is not taken into account in these equations.
 The contact term model with c=1 satisfies the phase theorem better than the pure VMD model due to the presence of the
contact term which increases significantly the magnitude the resonance term compared with the
background terms from the t and u channels.   This leads us to a smaller correction
 using the following integral equation approach where the unitarity relation is
explicitly built in. 

Eqs. (\ref{eq:3p1},\ref{eq:3p2},\ref{eq:3p3}) can be considereds as the large $N_c$ limit of the
QCD. This is true because $f_\pi^2 \sim N_c$ and the function $\Omega(s)$ defined by
Eq.(\ref{eq:vu1}) becomes in this limit a simple pole. We shall elaborate this fact later in this
article. 

In the following, we shall define  the function $G(s,t,u)$ at the symmetry point
$s=t=u=m_\pi^2$ while the chiral symmetry limit of this function is the chiral anomaly $\lambda$
given by Eq. (\ref{eq:anomaly3}). How are they related to each other? There is no clean answer for
this problem. Chiral Perturbation Theory could be used. The answer
depends, however, on one parameter
which is the scale parameter \cite{bijnens, hannah}. We prefer to look
at the large $N_c$ limit to get their relation.

 Setting $s=t=u=0$ in the
chiral limit in Eqs.(\ref{eq:3p1}, \ref{eq:3p2}), we have: 
\begin{equation}
G(s=m_\pi^2, t=m_\pi^2, u=m_\pi^2) \equiv \overline{\lambda}
=\lambda[1+\frac{3}{3-c}\frac{m_\pi^2}{m_\rho^2}] \label{eq:corr}
\end{equation}
This expression will be used in the following analysis. For c=1, we
have $\overline{\lambda}=1.049 \lambda $ whereas the corresponding
value for the one-loop ChPT \cite{bijnens, hannah}, assuming that the
scale parameter $\mu^2=m_\rho^2$, is $\overline{\lambda}=1.053 \lambda $
which is insignificantly larger.
In terms of $\overline{\lambda}$, with chiral symmetry broken but in the large $N_c$ limit, one has: 
\begin{equation}
G^{vmdc}(s,t,u)
=\frac{\overline{\lambda}}{3-c}\{\frac{m_\rho^2-m_\pi^2}{m_\rho^2-s}+\frac{m_\rho^2-m_\pi^2}{m_\rho^2-t}+
\frac{m_\rho^2-m_\pi^2}{m_\rho^2-u}-c\} \label{eq:3pb}
\end{equation}

\subsection{Vector Meson Model for  Pion Form Factor}
The solution of the integral for the pion form factor with the assumption of the elastic form factor
is
\begin{equation}
V(s) = P_n(s)\Omega(s) \label{eq:mo1}
\end{equation}
where $\Omega(s)$ is given by Eq.(\ref{eq:vu1}) and $P_n(s)$ is a
polynomial of degree n in s with real coefficients and $P_n(0)=1$.

For a given a set of the strong P-wave $\pi\pi$ phase shifts, the solution of the MO equation is not
unique. One can multiply the solution $\Omega(s)$ by a real polynomial to get a different set of
solutions with different asymptotic conditions. The low energy constraint enables us to fix at least
some coefficients of the polynomial.

 If one assumed $P_n(s)=1$, the square 
of the modulus of the pion form factor at the $\rho$ mass is too small by about 30\%, see Fig. 2, and the rms
radius of the pion is too small by 10\%. Constraining the rms radius to be equal to its experimental value, we
have to set \cite{truong4}
$P_n(s)=1+0.15(s/m_\rho^2)$ or:
\begin{equation}
V(s)=(1+0.15 \frac{s}{m_\rho^2}) \Omega(s) \label{eq:pff}
\end{equation}
The connection between this equation and the contact term was recently discussed \cite{truong3}.
The following integral equation for the process $\gamma\pi \rightarrow \pi\pi$ is more complicated
because the integral equation involves both the right and left cuts on the real axis and hence the
ambiguity of the solution is not simply the polynomial ambiguity and but is only related to it. It can
only be  obtained by solving numerically the integral equation as will be shown below.

\section{Integral Equation Approach using  Elastic Unitarity Relation }

In this article, the process $\gamma\pi \rightarrow \pi\pi$ is studied  
using dispersion relation and elastic unitarity for the lowest partial wave. A similar integral
equation of the type Muskhelishvilli-Omnes integral equation \cite{omnes} is
obtained.
 The difference is that the integral equation to be treated here is much more complicated due to 
crossing symmetry; no exact solution has been found. We shall get the solution of this 
integral equation by an iterative procedure, but with the crucial property that the 
iterative solution for the lowest partial wave satisfies the phase theorem at every steps as required
by unitarity \cite{watson}.  As the solution of the M-O equation is ambiguous by a polynomial, we
find here a similar problem. But the ambiguity is not the same i.e. a new solution cannot be obtained
by multipying the old solution by a polynomial.

We start first by deriving the single variables s,t,u  dispersion 
relation for $\gamma\pi \rightarrow \pi\pi$; we then project out the
P-wave amplitude where the rescattering effect is important due to presence of the low energy $\rho$
resonance at 0.77 GeV. The  rescattering effect is supposed to be
negligible for higher partial waves because there
are no resonances below 1.5 GeV for the two pions in F, H... waves. After solving the integral
equation numerically, one should put the results obtained in to a form of the single variables
dispersion relation. The crucial point is the single variable dispersion relation for the scattering
amplitude do not satisfy the phase theorem, but its P-wave projection does.

The integral equation can be derived using the technique of the Roy's equation for $\pi\pi \to
\pi\pi$ scattering \cite{roy}. We begin by writing a twice subtracted dispersion relation for
$G(s,t,u)$ at a fixed $t$. This
dispersion relation can be shown to be valid in general. Using the same technique as that used in
obtaining the Roy's equation, namely using the fixed t dispersion relation, crossing symmetry and
keeping only the P-wave for the partial wave expansion of the absorptive part, one arrives at ( for an
explicit demonstration of this equation see reference \cite{hannah}):
\begin{equation}
G(s,t,u)= \overline{\lambda} + [\frac{(s-m_\pi^2)^2}{\pi}\int_{4m_\pi^2}^\infty
\frac{\sigma(z) dz}{(z-m_\pi^2)^2(z-s-i\epsilon)} ] + [s \leftrightarrow t ]+ [s \leftrightarrow u] 
\label{eq:fstu} 
\end{equation}
where the symmetry point in the problem is at $s=t=u=m_\pi^2$, and $\overline{\lambda}$ is related
to $\lambda$ as will be discussed later. The subtracted linear terms do not contribute because they are proportional to 
$(s-m_\pi^2)+(t-m_\pi^2)+(u-m_\pi^2)=0$. One can make a partial fraction of the dispersion integral
to show that one can equally work well with the once subtracted dispersion relation which we shall
use in the following:
\begin{eqnarray}
G(s,t,u) & = & \overline{\lambda} + [\frac{(s-m_\pi^2)}{\pi}\int_{4m_\pi^2}^\infty
\frac{ImG_1(z) dz}{(z-m_\pi^2)(z-s-i\epsilon)} ] + [s \leftrightarrow t ]+ [s \leftrightarrow u]
\nonumber \\
     & = &  A(s) + A(t) + A(u) \nonumber \\
\label{eq:fstu1} 
\end{eqnarray}
where $ImG_1(z)$ is the imaginary part of the P-wave amplitude.

The assumption of the dominance of the P-wave amplitude to get the integral equation can be experimentally
checked by measuring the absence of the deviation from the $\sin^2\theta$ angular distribution given by Eq.
(\ref{eq:x}). Should this assumption be incorrect, one could try to treat phenomenologically the contribution
of the higher partial waves in the $\rho$ region by some real amplitudes \cite{hannah}.

Let us call $G_1(s)$ the lowest P partial wave projection of $G(s,t,u)$  as given by Eq.
(\ref{eq:pw2}).  The elastic
unitarity relation gives:
 \begin{equation}
ImG_1(s) = G_1(s)e^{-i\delta(s)} \sin\delta(s)  \label{eq:unitarity3}
\end{equation}
where $\delta $ is  the P-wave $\pi\pi$ phase shifts 
 obtained  from the available
experimental data which show  that they  pass 
through $90^{o}$ at the $\rho$ mass as can be seen from Fig. 1.
 There is no measurable inelastic effect below 1.2 GeV. Projecting out the P-wave from Eq.
(\ref{eq:fstu1}) and interchange the order of integration, we have:
\begin{eqnarray}
G_1(s)&=& \overline{\lambda} + \frac{s-m_\pi^2}{\pi}\int_{4m_\pi^2}^\infty
\frac{G_1(z)e^{-i\delta(z)}\sin\delta(z)}{(z-m_\pi^2)(z-s-i\epsilon)}dz  
 +\frac{3}{2\pi}\int_{4m_\pi^2}^\infty
G_1(z)e^{-i\delta(z)}\sin\delta(z) \times \nonumber \\
 & & \{
\frac{1}{b(s)}(1-\frac{(z-a(s))^2}{b(s)^2})\ln\mid\frac{z-a(s)+b(s)}{z-a(s)-b(s)}\mid+2\frac{
z-a(s)}{b(s)^2}-\frac{4}{3(z-m_\pi^2)}\}dz  \nonumber \\
\label{eq:int2}
\end{eqnarray}
The first derivative of $G_1(s)$ at $s=m_\pi^2$ vanishes and its second derivatives with
respect to s evaluated at $s=m_\pi^2$ is:
\begin{equation}
\frac{d^2 G_1(s)}{ds^2}\mid_{s=m_\pi^2}=\frac{12}{5\pi}\int_{4m_\pi^2}^\infty
\frac{ImG_1(z)dz}{(z-m_\pi^2)^3} \label{eq:deriv2}
\end{equation}

The standard solution of the M-O equation is ambiguous by a polynomial, but this problem is much  more complicated here
due to the symmetry of the s,t,u channels which leads to a much more complicated integral equation and
hence it does not have the same type of ambiguity.
Eq. (\ref{eq:int2}) is a complicated integral equation. It is similar to, but
more complicated than 
 the Muskelishvili-Omnes (MO) type \cite{omnes}, because the t and u channel contributions
 are also expressed in terms of the unknown function $G_1(s)$. It should be noticed that 
the first term has a cut from $4m_\pi^2$ to $\infty$ and the second one has a cut from
 $0$ to $-\infty$. For $s \geq 4m_\pi^2$ the argument of the  logarithm function in Eq.
 (\ref{eq:int2}) never vanishes and hence this enables one to solve the integral equation by
the following
 iteration scheme which converges very  fast.

\subsection {Iterative Solutions}

As it was remarked above, the integral equation (\ref{eq:int2}) has both right and left cuts. In
setting up the iterative scheme, it is important to keep in mind that the final solution should be
symmetric in the s,t,u variables as given by Eq. (\ref{eq:fstu1}). Because of this analytic structure,
we can define an iteration procedure
 which consists in splitting Eq. (\ref{eq:int2}) into two
 separate equations:

\begin{equation}
G_1^{(i)}(s) = \frac{ \overline{\lambda} }{3} + T_B^{(i-1)}(s) + 
\frac{s-m_\pi^2}{\pi}\int_{4m_\pi^2}^\infty
\frac{G_1^{(i)}(z)e^{-i\delta(z)}\sin\delta(z)}{(z-m_\pi^2)(z-s-i\epsilon)}dz  \label{eq:inti}
\end{equation}
and

\begin{eqnarray}
T_B^{(i-1)}(s) &=& \frac{2\overline{\lambda}}{3}+\frac{3}{2\pi}\int_{4m_\pi^2}^\infty
G_1^{(i)}(z)e^{-i\delta(z)}\sin\delta(z) \times \nonumber \\ 
 & & \{\frac{1}{b(s)}(1-\frac{(z-a(s))^2}{b(s)^2})\ln\mid\frac{z-a(s)+b(s)}{z-a(s)-b(s)}\mid+2\frac{
z-a(s)}{b(s)^2}-\frac{4}{3(z-m_\pi^2)}\}dz  \nonumber \\
  \label{eq:tbi-1}
\end{eqnarray}
 where $i\geq 1$ and $G_1^{(i)}$ is the value of the function $G_1(s)$ calculated at the $i^{th}$
step
 in the iteration procedure; the Born term $ T_B^{i-1}(s)$ is calculated at the $i^{th}-1$ 
step. An iteration cycle is defined as a numerical calculation of both these two equations.

 The Born term is real for $s\geq 0$ and has a left cut in s for $s<0$.
 In writing Eqs. (\ref{eq:inti},\ref{eq:tbi-1}),
 care was taken to preserve the symmetry in the $s,t,u$ variables 
for the function $F(s,t,u)$ which requires us to 
split symmetrically the subtraction constant $\overline{\lambda}$ in Eq. (\ref{eq:int2})
 into three equal pieces, one contributes to Eq. (\ref{eq:inti}) the other two to
Eq. (\ref{eq:tbi-1}). 

The solution of the integral equation Eq. (\ref{eq:inti}) is of the MO type \cite{omnes}:

\begin{equation}
G_1^{(i)}(s)= \frac{\overline{\lambda}}{3}\overline{\Omega}(s,m_\pi^2) + T_B^{(i-1)}(s) +
 \overline{\Omega}(s,m_\pi^2)\frac{s-m_\pi^2}{\pi}
\int_{4m_\pi^2}^\infty \frac{\overline{\Omega}^{-1}(z,m_\pi^2) e^{i\delta(z)}\sin\delta(z)
T_B^{(i-1)}(z) dz} {(z-m_\pi^2)(z-s-i\epsilon)} \label{eq:sol1}
\end{equation}
where:
\begin{equation} 
\overline{\Omega}(s,m_\pi^2) =  \frac{P_n(s)\Omega(s)}{P_n(m_\pi^2)\Omega(m_\pi^2)} \label{eq:omeb}
\end{equation}
i.e. this new function $\overline\Omega$ is normalized to unity at $s=m_\pi^2$ and $P_n(s)$ are
polynomial of nth degree with real coefficients. In the following,
similar to Eq. (\ref{eq:pff}), we
only take the first to terms in the polynomial and hence set:
\begin{equation}
P_n(s) = 1+\alpha \frac{s}{m_\rho^2} \label{eq:poly}
\end{equation}
where $\alpha$ is a parameter which is related to the contact term $c$ defined previously. The second
derivative of the P-wave amplitude, defined by the sum rule Eq. (\ref{eq:deriv2}), depends sensitively 
on the parameter $\alpha$.

 Eq. (\ref{eq:sol1}) is not really typical solution written down for this type
of  integral equation. It is usually  written in terms of the "driving" term $T_B^{(i-1)}(z)$. This procedure  is not at
all valid for the present situation, but we must modify it in order to get the final solution for
the full amplitude with is completely symmetric in the s, t, u variables.  Eq. (\ref{eq:sol1}) is
written  with this fact in mind.  The first term on its RHS  represents the VMD in the s-channel with
or without contact term, the second term is the corresponding contribution from the t and u channels
and the third term is the rescattering due to the final state interaction in the s-channel.

At first sight one would think that the RHS of Eq. (\ref{eq:sol1}) does not have the P-wave phase
$\delta$. This is not so because we first note that  the last integral can be separated into a 
principal part integral and a delta-function contribution which is purely imaginary. Then combining
this delta-function contribution with
the $T_B^{(i-1)}(s)$ in Eq.(\ref{eq:sol1}), we have:
\begin{eqnarray}
G_1^{(i)}(s) &=& \overline{\Omega}(s,m_\pi^2)\{\frac{\overline{\lambda}}{3} +
T_B^{(i-1)}(s)Re[\overline{\Omega}^{-1}(s,m_\pi^2)]+ \nonumber \\ 
& &\frac{s-m_\pi^2}{\pi}P
\int_{4m_\pi^2}^\infty \frac{\overline{\Omega}^{-1}(z,m_\pi^2) e^{i\delta(z)}\sin\delta(z)
T_B^{(i-1)}(z) dz} {(z-m_\pi^2)(z-s)} \} \nonumber \\
\label{eq:sol2}
\end{eqnarray}
where P stands for the principal part integration. We have made the
usual decomposition N/D for the partial wave amplitude, which can be
shown to be quite general and independent of any dynamical scheme: 
$e^{i\delta}\sin\delta(s)= N(s)\rho(s)\overline{\Omega}(s,m_\pi^2)$ where $N(s)$ contains only the
left hand cut,$ D^{-1}(s)=\overline{\Omega(s)}$ only the right hand cut and 
$\rho(s)=\sqrt{(1-4m_\pi^2/s)}$. Eq.(\ref{eq:sol2}) shows indeed $G_1^{(i)}(s)$ has the phase $\delta$.

 One  arbitrarily defines the convergence of the iteration scheme at the ith iteration step 
when $\mid G_1^{(i)}\mid/\mid G_1^{(i-1)}\mid$ differs from 1 by less than 1\% or so in the
energy range
 from the two pion threshold to $1 GeV$. (Alternatively one can also require that the ratio 
$\mid T_B^{(i)}\mid/\mid T_B^{(i-1)}\mid$ to be unity within an accuracy of 1\%).

Once the solution for the partial wave is obtained, one should return to
the calculation of the full amplitude. 
 This can be done by 
combining the
$T_B^{(i-1)}$ Born term in Eq. (\ref{eq:sol1})
 with higher uncorrected 
partial waves (for rescattering) from the $t$ and $u$ channels to get the final solution:
\begin{equation}
G^{(i)}(s,t,u)= \frac{\overline{\lambda}}{3}[
\{\overline{\Omega}(s,m_\pi^2)(1+3I^{(i-1)}(s))\} +
\{(s\leftrightarrow t)\}
 +\{(s\leftrightarrow u)\} ]
 \label{eq:final}
\end{equation}
where the function $I^{(i-1)}$ denotes the multiple rescattering correction:
\begin{equation}
I^{(i-1)}(s) = \frac{s-m_\pi^2}{\pi} \int_{4m_\pi^2}^\infty 
 \frac{\overline{\Omega}^{-1}(z,m_\pi^2) e^{i\delta(z)}sin\delta(z) T_B^{i-1}(z) dz}
{(z-m_\pi^2)(z-s-i\epsilon)} \label{eq:I}
\end{equation}
It is obvious that $G^{(i)}(s,t,u)$ do not have the phase $\delta$ of the P-wave $\pi\pi$ scattering,
but its P-wave projection does. This is so because,  projecting out the $l=1$ partial wave from Eq.
(\ref{eq:final}), we arrive at Eq. (\ref{eq:sol1})
 with $T_B^{i-1}(s)$ replaced by $T_B^{i}(s)$.
Because of
 the assumed criteria for the convergence of the iteration scheme, $T_B^{i-1}(s)\simeq T_B^{i}(s)$,  
 it is easily seen  that 
$G_1^{(i)}(s)$ has the phase $\delta$, using the result of Eq. (\ref{eq:sol1}) and
Eq. (\ref{eq:sol2}).
 The remaining  higher partial waves $l>1$
 are all real because we
 have assumed that the strong final state interaction 
 of the higher partial waves are  negligible.
 The final solution Eq. (\ref{eq:final}) is completely symmetric in the $s,t,u$ variables.

\section{ Numerical Solutions}
We shall solve numerically our integral equation for various values of $\alpha$ defined by
Eq. (\ref{eq:poly}), corresponding to different values of the contact terms as discussed in
Eq. (\ref{eq:3p2}). We examine the following case: $\alpha=0, 0.30, 0.50, 0.70$. The iteration
scheme can be done by first guessing a solution for
$T_B(s)$ corresponding to a chosen value of $\alpha$. We can take a rather arbitrary first solution
for this function. For example, for a given $\alpha$ we can take the t and u channel contribution to
the Born term  as:
\begin{equation}
T_B^{0}(s)= \frac{3}{4}\int_{-1}^1 d\cos\theta \sin^2\theta
\frac{\overline{\lambda}}{3}\{[\frac{m_\rho^2-m_\pi^2}{m_\rho^2-t}]+
[\frac{m_\rho^2-m_\pi^2}{m_\rho^2-u}] \}\label{eq:born1}
\end{equation}
which is independent of $\alpha$. With this expression for the Born term, the iteration
scheme can be started by calculating the solution of the integral equation as given by
Eq. (\ref{eq:sol1}). The next step is to calculate the Born term by Eq. (\ref{eq:tbi-1}) where there
is no arbitrariness at this step. The iteration scheme is continued until the convergence of the
solution is obtained.

The number of the iterations depend on the original choice of the Born term i.e. how closed  it is to
the final solution. Even for a not so good  approximation for $T_B^{(0)}$ 
as given by Eq. (\ref{eq:born1}), it is found that after one iteration one can already reach a
reasonable approximation for the solution of the integral equation.

 Convergence to the final solution  for
$\alpha= 0.3, 0.5, 0.7$,
with a precision of the order of 1\% or better, could be achieved without
using the iteration scheme if one chose a good expression for $T_B^{0}(s)$. For this purpose, one  used instead  the Born terms calculated from the following zero width
contribution  of the t and u channels  of Eq. (\ref{eq:3p2}) which
depends on the strength of the contact term:
\begin{equation}
T_B^{0}(s)= \frac{3}{4}\int_{-1}^1 d\cos\theta \sin^2\theta
\frac{\overline{\lambda}(1+\alpha)}{3}\{[\frac{m_\rho^2-m_\pi^2}{m_\rho^2-t}]+
[\frac{m_\rho^2-m_\pi^2}{m_\rho^2-u}]-\frac{2\alpha}{1+\alpha}\}\label{eq:born2}
\end{equation}
where we have used the large $N_c$ relation: 
\begin{equation}
\alpha = \frac{c}{3-c} \label{alpha}
\end{equation}
 We only make use of this relation here
to calculate the Born terms and also to get the relation between
$\lambda$ and $\overline{\lambda}$.

This relation is obtained in the limit of of a narrow $\rho$ width.  As it was discussed above, this limit is obtained when we let
the number of color $N_c \to \infty$. In this limit, as it will be shown below, Eq. (\ref{eq:final})
becomes Eq. (\ref{eq:3p2}) or Eq. (\ref{eq:3p3}). The pure
VMD model corresponds to
$\alpha=0$ and for models with the contact term c=1, e.g.  the hidden
symmetry model \cite{bando}, $\alpha=1/2$. (More precisely, the hidden symmetry model with no contact term in
the pseudoscalar mesons decaying into 2 gammas requires  c=1). For the real
situation where 
$N_c=3$ there should be substantial correction to this relation.

For
$\alpha=0$, without an iteration of the integral equation, one can only get a precision of the order of
5\%. Similarly for other values of $\alpha$, using Eq. (\ref{eq:born2}) for the
Born terms, and without going through the iteration scheme, one can already achieve a precision of
better than 1\% using only Eq. (\ref{eq:inti}).

For $\alpha=0$, after 5 iterations a precision of better than 1\% is reached. For other values
of $\alpha$= 0.3, 0.5, 0.7 a precision of better than $10^{-3}\%$ is reached after only 4 iterations.
These numbers indicate the rescattering effect is much more important for $\alpha=0$ and is much less
important for other values of $\alpha$.  

The slow convergence of the iteration scheme for $\alpha=0$ is due to a large violation of the phase
theorem at the zeroth order as shown in Fig. 3. For other cases the violation of the phase theorem is
not so serious and even without the iteration scheme one can already get the approximate solution
accurate to better than 1\% by solving directly the integral equation
as discussed above.

 Instead of parametrizing our solution by the
value of $\alpha$, it is more physical to describe the solution as a function of the width
$\Gamma(\rho \to \pi\gamma)$.  This quantity is not unambiguous and
will be defined in the following section. It is denoted by
$\Gamma(\rho \to \pi\gamma)$ using our definition while the corresponding partial
width using the usual  Breit-Wigner parametrization is denoted by
$\Gamma(\rho \to \pi\gamma)_{bw}$ with the value of the $\rho$ mass the same as our defintion, i.e
$m_\rho^2=0.593 GeV^2$ and $\Gamma (\rho \to \pi\pi)=0.156 GeV$. There is a substantial difference for the
values of 
$\Gamma (\rho
\to \pi\gamma)$  using these two definitions. In Table 1. $\Gamma (\rho \to \pi\gamma)$, $\Gamma(\rho \to
\pi\gamma)_{bw}$ and the second derivative of the P-wave amplitude at
$s=m_\pi^2$ are given as a function of the numerical value of $\alpha$.

  Corresponding to various
values of $\alpha$ the square of the absolute values of the P-wave
amplitudes $G_1(s)$ are plotted against the energy square s, in the unit of
$\overline{\lambda}$, are given in Fig. 4 and Fig. 5. It is seen that
the larger the values of $\alpha$ the higher is the maximum values of
the P-wave amplitudes. At lower energy, the absolute value of the
magnitude of the P-wave amplitude also increases with $\alpha$ until
$\alpha=0.5$ and then decreases with larger values as shown in Fig. 5. 

In Fig. 6 the modulus of the ratio $3G_1(s)/\overline{\Omega}(s)$ is plotted against the energy squared s(GeV$^2$); this ratio indicates the deviation from the Breit-Wigner form  as given by the function $\overline{\Omega}(s)$.

For various values of $\alpha$, accurate values (to better than 1\%) of the modulus of the P-wave
amplitude from the two pion threshold to 1 GeV can be obtained by using the modulus of $\overline{\Omega}(s)$ divide the function  C(s)
given in Table 2:
\begin{equation}
G_1(s) = \overline{\Omega}(s, m_\pi^2) C(s) \label{eq:C}
\end{equation}

We  are also interested in finding out what are the corrections due to the multiscattering effect in
the VMD approximation to the function A(s) defined by Eq. (\ref{eq:fstu1}) and given by Eq.
(\ref{eq:final}): 
\begin{equation}
A(s) = \frac{\overline{\Omega}(s,m_\pi^2)}{3} J(s) \label{eq:ff2}
\end{equation}
or
\begin{equation}
J(s) =\frac{\lambda}{3} ( 1 + 3I^{i-1}(s) ) \label{eq:ff3}
\end{equation}
for a value of i attained at the end of the iteration of the integral equation.
In Fig. 7 the modulus of J(s) is plotted against the energy squared s
in the unit of GeV$^2$.  If there were no
corrections to the VMD model, J(s) would be unity. It is seen that the corrections are most
important for the case of $\alpha=0$. 

 In Fig.
8 the phase of A(s) is also plotted against  the energy squared s  for various values of $\alpha$
and compared with the P-wave $\pi\pi$ phase shifts.

In Fig. 9, $\mid G(s, t, u)\mid^2$ with $\cos\theta=0$ and $\cos\theta=0.75$   are plotted
against s (in GeV$^2$) are shown for the special case
$\alpha=0.5$. Figs. with other values of  $\alpha$ and $\cos\theta$
are not shown because they are quite similar to the Fig. 9. Therefore the
higher partial waves are completely negligible for an energy below 1 GeV.

\section{Comparison with experimental data and other theoretical
  works}

Our calculation can be compared with the experimental data at low and
high energy. At low energy, the only experimental data available is
given by reference \cite{antipov}. From Fig. 5, at an energy $s \simeq 0.16 GeV^2$, corresponding to the average
energy measured in the reference \cite{antipov} we have, for
  $\alpha$=0.5, $\mid G_1(s) \mid \simeq \mid G(s,t=u) \mid 
  =1.15 \lambda $ which is about one standard deviation  smaller than the measured value 
  $(1.29\pm.09\pm .05) \lambda$. It is clearly important to improve the
  precision of this experiment.

At higher energy, the experimental cross section for  $\gamma \pi \to \pi\pi$ are usually analyzed in terms of the
Breit Wigner formula:
\begin{equation}
\frac{d\sigma}{ds} = \frac{24\pi s}{(s-m_\pi^2)^2} \frac{m_\rho^2 \Gamma(\rho\to
2\pi) \Gamma(\rho\to\pi\gamma)}{(m_\rho^2-s)^2+m_\rho^2 \Gamma_\rho^2(s)} \label{eq:bw}
\end{equation}
This formula is usually not accurate because it either neglects the contribution of the part of the
amplitude from the t and u channels or assumes that the cross section can be fitted with a Breit
Wigner form which may not be true. Furthermore the maximum of the modulus of the P-wave amplitude is
shifted significantly toward lower energy which complicates  the analysis of the
experimental data using Eq. (\ref{eq:bw}).

The result of our calculation shows that, at the maximum of the absolute
 value of the P-wave amplitude, the phase of the amplitude is not 90$^\circ$.
 The only reasonable method which we can find acceptable
is to define the $\rho$ mass as the value of s when the phase of the function $\Omega$, which is the
same as the experimental P-wave $\pi\pi$ phase shift passing through 90$^\circ$. Its width is
proportional to the  inverse of of the derivative with respect to s of the $\cot\delta$ at
$s=m_\rho^2$:
\begin{equation}
\frac{1}{m_\rho \Gamma_\rho} = \frac{d}{ds}\cot\delta(s) \mid_{s=m_\rho^2} \label{width}
\end{equation}
With this definition the $\rho$ width as given by Eq.(\ref{eq:vu1}) is 0.156 GeV. 
One then could use the P-wave cross section at $s=m_\rho^2$ to calculate the $\Gamma(\rho \to
\pi\gamma)$ width using Eq. (\ref{eq:bw}). The values obtained is
denoted by  $\Gamma (\rho \to \pi\gamma)$ and  would approximately be 10\% lower than the value
obtained  from using the maximum observed
$\gamma
\pi \to \pi\pi$ cross section in combination with the Breit-Wigner
formula, Eq. (\ref{eq:bw}) which is now  denoted by  $\Gamma (\rho \to
\pi\gamma)_{bw}$ (see Table 1).

 Using our method, we could even integrating the measured cross section
on either side of the $\rho$ mass by 0.1 GeV in order to improve the experimental accuracy without
changing its value by more than 1\%. This precision would not be possible if one did the calculation
with the $\rho$ mass as the value of the maximum cross section.

The present experimental results are not consistent with each other. The more recent published
experimental results by Caparo \emph{et. al}. gave the value for $\Gamma(\rho \to \pi\gamma)=81\pm4\pm4$ KeV
\cite{capraro}, wheras earlier results by Huston \emph{et. al}. gave a lower value \cite{huston}. These two
experiments are the Primakoff-like experiments using a high energy charge pion beam on a heavy target.
 The experimental result
from
 $e^+e^-$ reaction gives a higher value for the $\rho \to \pi\gamma$ width \cite{dolinsky} but
experimental results contain large error.

A more recent unpublished result using photo-production of a pair of pions off the nucleon target yields 
  $\Gamma (\rho \to \pi\gamma)_{bw}=96 \pm12 KeV$ \cite{bernstein}. Unlike two previous Primakoff
experiments, this experiment might have some difficulties with isolating the data corresponding to the one
pion exchanged diagram from the background effect; one must also take into account of the fact that  the
exchanged pion is off its mass shell.

 Due to the lack of  experimental informations on the
second derivative of the P-wave amplitude  at s=0 or the
parameter $\alpha$ (see the Table 1), we cannot predict  the solution of the integral equation to get the
$\Gamma(\rho
\to
 \pi\gamma)$ width.

 Corresponding to a naive  pure VMD model without a contact term, from Table 1, our calculation with
$\alpha=0$  yields a width of
$\Gamma (\rho
\to
\pi\gamma)= 50.7 KeV$  or $\Gamma (\rho \to
\pi\gamma)_{bw}=57.8 KeV$, whereas corresponding to the hidden symmetry model with c=1, our calculation with
$\alpha=0.5$ yields
$\Gamma (\rho \to\pi\gamma)= 84.3 KeV$ or $\Gamma (\rho \to
\pi\gamma)_{bw}=91.8 KeV$.

With $\alpha=0.5$ the value for $\Gamma (\rho \to
\pi\gamma)_{bw}$ is somewhat smaller than the value of
96 KeV obtained
 by Hannah  using
the Pade or the inverse amplitude methods for ChPT two loop
 amplitude which was calculated numerically \cite{hannah}. For this special value of $\alpha$,
 one would also recover the main
result of the hidden symmetry  model with c=1. The low
energy parameters $\overline{C}$ and $\overline{D}$, corresponding to
 the first and second derivatives of the function A(s), Eq. (\ref{eq:ff2}), defined and
evaluated by Hannah \cite{hannah} are in agreement to an accuracy of 2\%
with those from the our integral equation approach.

 The difference between this work with that of Hannah is presumably due to the interpretation of the
Eq. (\ref{eq:bw}), the treatment of the multiple scattering effects and also to the
interpretation of the contact term.  Hannah's work shows the importance of the resummation of the
perturbation series by the inverse amplitude  or the Pade methods. 

There is a similar treatment of this problem by Holstein. Holstein solution was
obtained by taking the product of the three functions:
\begin{equation}
G_1^{H}(s,t,u) = \lambda P_n(s,t,u) \Omega(s)\Omega(t)\Omega(u) \label{eq:holstein}
\end{equation}
where $P_n(s, t, u)$ is a polynomial in s, t, u constructed in such a
way that this equation has the same low energy limit as that given by ChPT.
The merit of this equation is the phase theorem is explicitly obeyed as can be seen by projecting
out the P wave from this equation. But this equation is not right because all higher partial
waves such as F, H ...all have the $\rho$ resonance or they all have the phase of the P-wave phase
shifts which are not correct.
 The singularity associated with the multiple scattering effects which are present in
our integral equation approach do not contain in Eq.(\ref{eq:holstein}). All possible solutions which
can be written in terms of the product form of three  functions in s,t,u variables will have
this problem. One  exception is the problem involving three hadrons with two light
particles having no interactions between them but they interact with an infinitely heavy target.

Holstein's solution yields a comparable value for $\Gamma (\rho \to\pi\gamma)_{bw}$ as our solution with
$\alpha=0.5$. This result is not surprising because the value of the second derivative  of his solution at 
$s=m_\pi^2$ is also comparable with ours. His solution can be fixed by projecting out the P-wave imaginary
part
 and put it in Eq. (\ref{eq:fstu}) to provide the necessary corrections.

It should be reemphasized that our result  is not in the product form as in
Eq.(\ref{eq:holstein}) but as a sum of the three identical functions with the interchange of s,t,u
variables Eq.(\ref{eq:fstu}) and Eq.(\ref{eq:final}). It is a direct consequence of the fixed t
dispersion relation, using crossing symmetry and neglecting the contribution form the higher partial
waves at low energy in the absorptive parts.

\section {Importance of the Multiple Scattering Correction as a Function of the $\rho$ Width}

Our formulation of the problem $\gamma \pi \to \pi\pi$ is quite useful in understanding  the importance
of the multiple scattering effect as a function of the $\rho$ width. We have previouly stated that in
the large N$_c$ limit the multiple scattering effect should vanish and we should recover the VMD models
as given by Eqs. (\ref{eq:3p1},\ref{eq:3p2},\ref{eq:3p3}). In order to
see that this statement is correct,
it is sufficient to study the correction factor J(s) defined by Eq. (\ref{eq:ff2}) as a function of
the $\rho$ width. It is sufficient to study this question for the case
when $\alpha=0$. In Fig. 10, the
modulus of the function J(s) is plotted against the energy squared when the
$\rho$ widths are increased or decreased by a factor of 4 when $f_\pi$ is changed by a factor of 2.
This can be seen from examining the definition of the function of
$\Omega(s)$, Eq. (\ref{eq:vu1}). The result is the multiple scattering effect increases as the
$\rho$ width increases, and decreases as the the
$\rho$ width decreases. It is easy to see that for a zero width resonance, the correction factor J(s) is
unity. 

The result of this section provides also some arguments for the neglect of the multiple scattering
effect in the study of $<\gamma\mid 3\pi>$ and $<3\pi \mid \pi\gamma>$ with the 3$\pi$ resonating as
the
$\omega$ state because of the extreme small width of this resonance
\cite{truong3}.

\section{Conclusion and Acknowledgment}

We have studied in this article the scattering of $\gamma \pi \to
\pi\pi$ using the integral equation approach. Because the second
derivative of the P-wave amplitude at $s=m_\pi^2$ is not known, the
maximum cross section for this process cannot be predicted with
reliability. The solution of the integral equation is ambiguous and
depends on the second derivative of the P-wave amplitude at
$s=m_\pi^2$. This problem is  similar to problem of the contact term in the usual
VMD model.

If the ambiguity of the solution of the integral equation could be
interpreted as the imperfectness  of the elastic unitarity relation in
describing the low energy phenomena, then
one would have to be satisfied with a precision of the order of 15\% in
amplitude for the pion form factor calculation \cite{truong1}. This
inadequacy could then be removed using the knowledge of the pion
r.m.s. radius \cite{truong1}.

 For the  $\gamma \pi \to
\pi\pi$ calculation, this inadequacy  could become more
serious  due to the existence of 
singularities associated with  the t and u channels. Furthermore, the
corresponding first derivative of the P-wave amplitude vanishes due to
the symmetry of the problem, and hence we could only use the knowledge
of the second derivative at $s=m_\pi^2$ to fix up the inadequacy of
the elastic unitarity relation. This last parameter could not be
precisely measured and hence we could not predict with certainty 
$\Gamma(\rho\to\pi\gamma)$.

We show in this article that  there is a one to one correspondance between the
contact term in the VMD model and the ambiguity
associated with the parameter $\alpha$ in our
integral equation approach. This parameter plays the role of the
second derivative of the P-wave amplitude at $s=m_\pi^2$. 

It is a pleasure to thank Torben Hannah for sending me a copy of his
paper before publication and for useful discussions.

\newpage
\begin{center}
\vspace{5mm}

\begin{tabular}{|c|c|c|c|} \hline
$\alpha$ & $\Gamma(\rho \to \pi\gamma)$ KeV & $\Gamma(\rho \to
\pi\gamma)_{bw} $ KeV & $\overline{\lambda}^{-1}d^2G_1(s)/ds^2(s=m_\pi^2)
GeV^{-2}$
\\
\hline 0.00 & 50.7 & 57.8 & 4.88 \\ \hline 
0.30 & 68.0 & 76.4 & 5.30  \\ \hline
0.50 & 84.3 & 91.8 & 5.64 \\ \hline
0.70 & 103 & 111 & 6.00 \\ \hline
\end{tabular}

\vspace{2mm}

Table 1:Solution of the  P-wave amplitude for the $\gamma \pi \to \pi \pi$ integral equation as a
function of the parameter $\alpha$. The second column is the
$\Gamma(\rho \to \pi\gamma)$ in KeV according to the definition given
in the text. The third column is the  $\Gamma(\rho \to \pi\gamma)_{bw}$ in
KeV using the Breit Wigner formula Eq. (\ref{eq:bw}) evaluated at the maximum of
  the cross section;
 The fourth column is the second derivatives of the P-wave amplitude at the point $s=m_\pi^2$
in the $GeV^{-2}$ unit.

\end{center}

\bigskip
\begin{center}
\vspace{5mm}

\begin{tabular}{|c|c|} \hline
 $\alpha$  & $\frac{1}{C(s)}$ \\ \hline
 0.00 & $ 2.457s^3-1.662s^2+1.162s+0.2813 $ \\ \hline
0.30 & $ 0.808s^2+0.667s+0.3250 $ \\ \hline
0.50 & $ -0.759s^3+1.278s^2+0.705s+0.3267 $ \\ \hline
0.70 & $ -1.015s^3+1.183s^2+0.898s+0.318 $ \\ \hline
\end{tabular}

\vspace{2mm}

Table 2: Relation between the P-wave amplitude $G_1(s)$ and $\overline{\Omega}(s,m_\pi^2)$ as given
by the function C(s) defined by Eq. (\ref{eq:C}) given in text.

\end{center}

\newpage

{\bf Figure Captions}

Fig.1~: The phases of the function $\Omega(s)$  in degrees (vertical axis) are given as a function of
$s(GeV^2)$. The experimental data are taken from references
\cite{proto, hyams,  martin}.

Fig.2~: The square of the modulus of the function $\Omega(s)$
(vertical axis) is given as a function of $s$ in
$GeV^2$ (dashed line). The square of the modulus of the pion form factor $V(s)$ as given by Eq.
(\ref{eq:pff}) (solid line). Experimental data are taken from
references \cite{barkov, aleph}.

Fig.3~: The P-wave strong $\pi\pi$ phase shifts (vertical axis) as a function of the
energy  are shown by the solid line. The projected P-wave amplitude phases
of the $\gamma \pi^0 \to \pi^+ \pi^-$ as given by the VMD model
without contact term, Eq. (\ref{eq:3p11}), are given by the long dashed
line;
with the contact term c=1,
Eq. (\ref{eq:3p22}), are given by the short  dashed line; 
 with the
contact term c=1,
Eq. (\ref{eq:3p33}), are given by the medium dashed line.

Fig.4~: Plot of the square of the absolute value of the P-wave
amplitude $\mid G_1(s)\mid^2$ in the unit of $\overline{\lambda}^2$ as a function of s in $GeV^2$
for $\alpha=0.5$ (solid line); 
$\alpha=0.0$ (short dashed line); $\alpha=0.3$ (medium dashed line); 
$\alpha=0.7$ (long dashed line).

Fig.5~: Same as Fig. 5 but with $0.08 GeV^2 \leq s \leq 0.25 GeV^2$.

Fig.6~: Plot of the ratio $3G_1(s)/\Omega(s,m_\pi^2) $ in the unit of $\overline{\lambda}$ for
various values of $\alpha$. The curves are same as in Fig.4.

Fig.7~: Plot of the absolute value of the function J(s) defined by Eq.
(\ref{eq:ff2}) in the unit of $\overline{\lambda}$ as a function of s in $GeV^2$.   The curves are same as in Fig.4.

Fig.8~: Plot of the phases of the function A(s) 
 defined by Eq.
(\ref{eq:ff2}) as a function of s in $GeV^2$. The solid line
 represents the P-wave strong $\pi\pi$ phase shifts; the dotted line,
 $\alpha=0.5$; the short dashed line, $\alpha=0.0$; the medium dashed,
 $\alpha=0.3$, the long dashed line, $\alpha=0.7$.

Fig.9~: Square of the modulus of G(s,t,u) in the unit of
$\overline{\lambda}^2$ with $\cos\theta=0$ (long dashed line), with
$\cos\theta=0.7$ (short dashed line) and the square of the modulus of the P-wave
amplitude $G_1(s)$ (solid line) for
$\alpha=0.5$  as a
function of s in $GeV^2$.

Fig.10.~: Plot of the absolute value of J(s)(vertical axis)  defined by Eq.
(\ref{eq:ff2}) in the unit of
$\overline{\lambda}$  vs s (GeV$^2$) for $\alpha=0$ for various values of the $\rho$ widths. 
 The solid curve represents $\Gamma_\rho=0.156 GeV$, the short dashed 
curve, $\Gamma_\rho=0.039 GeV$, the long dashed
curve, $\Gamma_\rho=0.624 GeV$.

\newpage
\begin{figure}
\def\picsize{9cm}
\epsfxsize \picsize
\epsfbox{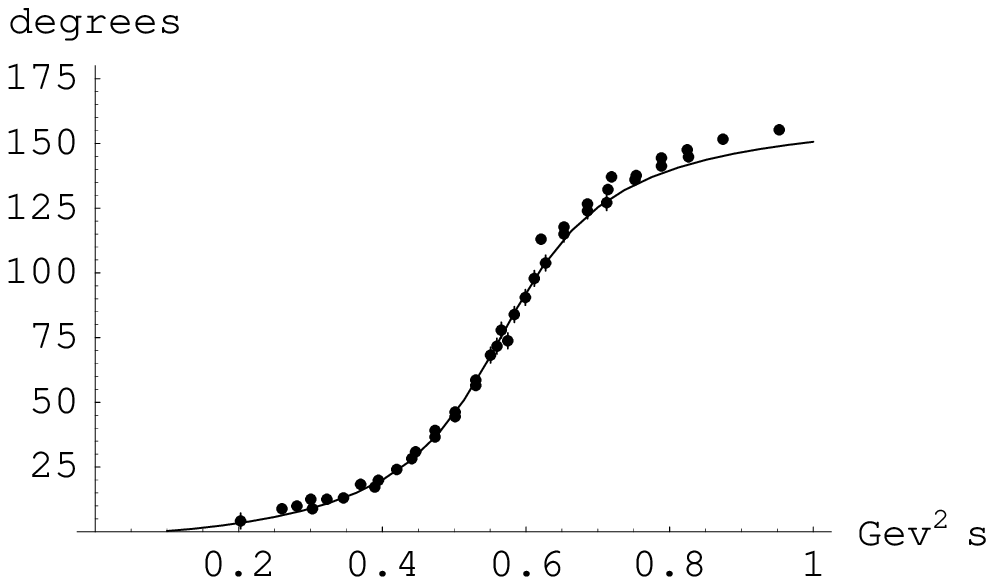}
\caption{}
\label{Fig.1}
\end{figure}

\begin{figure}
\def\picsize{9cm}
\epsfxsize \picsize
\epsfbox{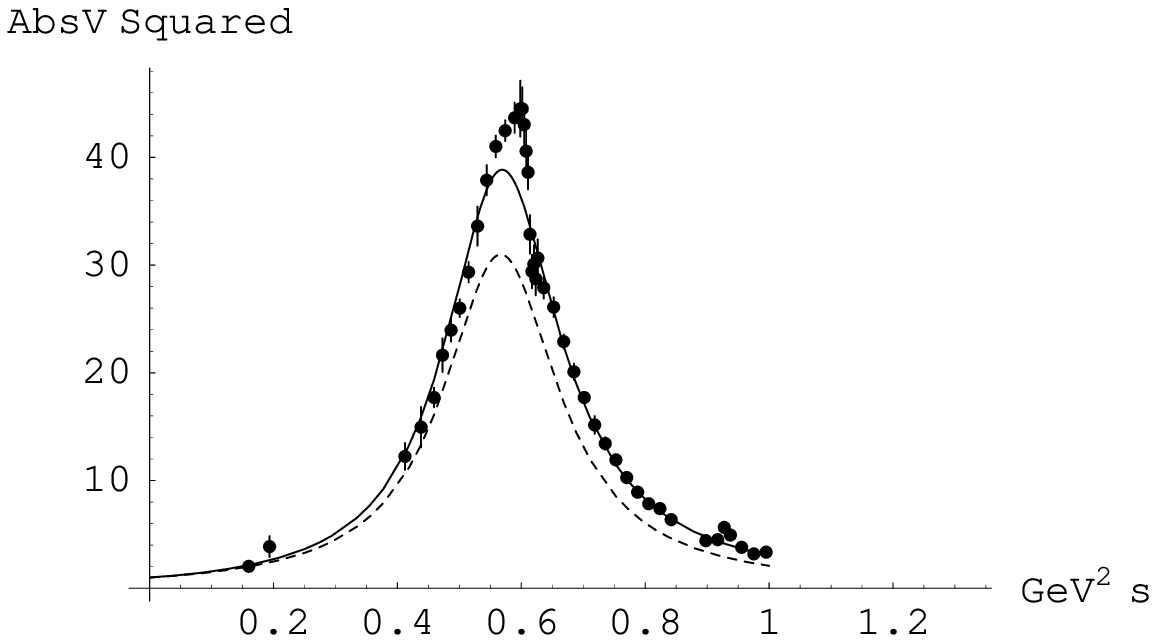}
\caption{}
\label{Fig.2}
\end{figure}
\vspace{5cm}

\newpage

\begin{figure}
\def\picsize{9cm}
\epsfxsize \picsize
\epsfbox{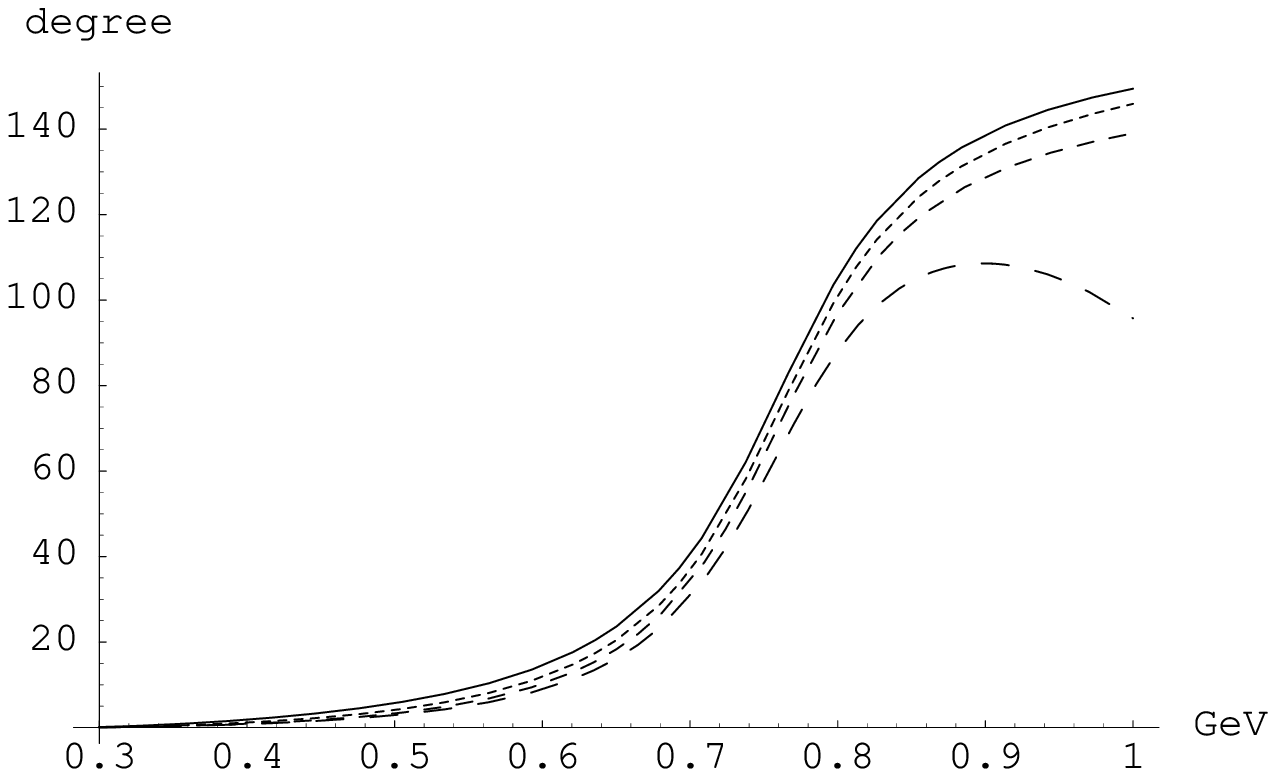}
\caption{}
  \label{Fig.3}
\end{figure}

\begin{figure}
\def\picsize{9cm}
\epsfxsize \picsize
\epsfbox{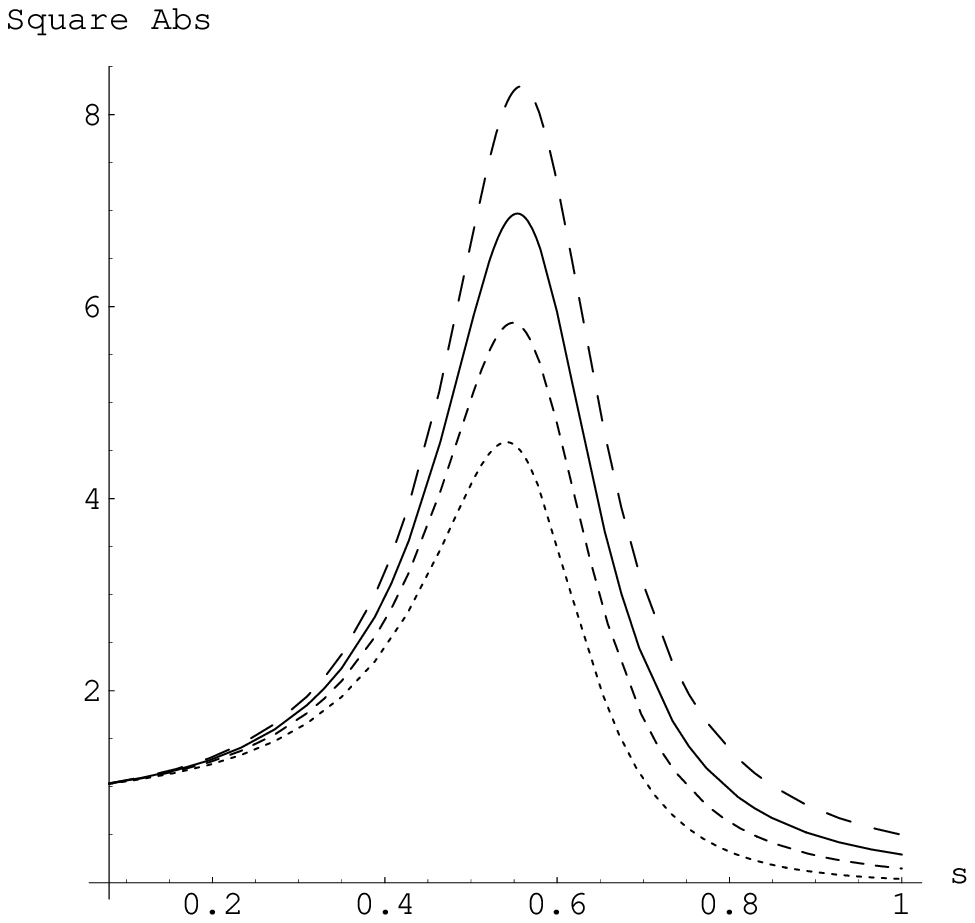}
\caption{}
\label{Fig.4}
\end{figure}

\begin{figure}
\def\picsize{9cm}
\epsfxsize \picsize
\epsfbox{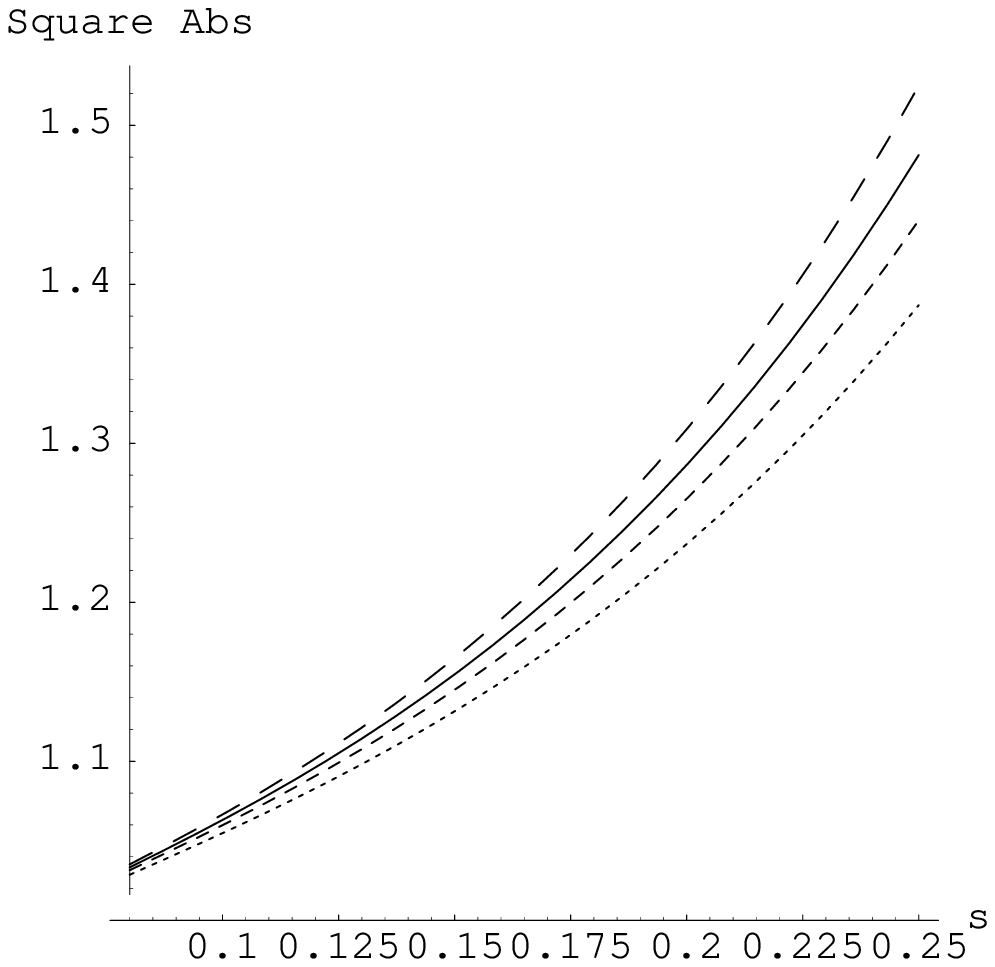}
\caption{}
\label{Fig.5}
\end{figure} 

\newpage
\begin{figure}
\def\picsize{9cm}
\epsfxsize \picsize
\epsfbox{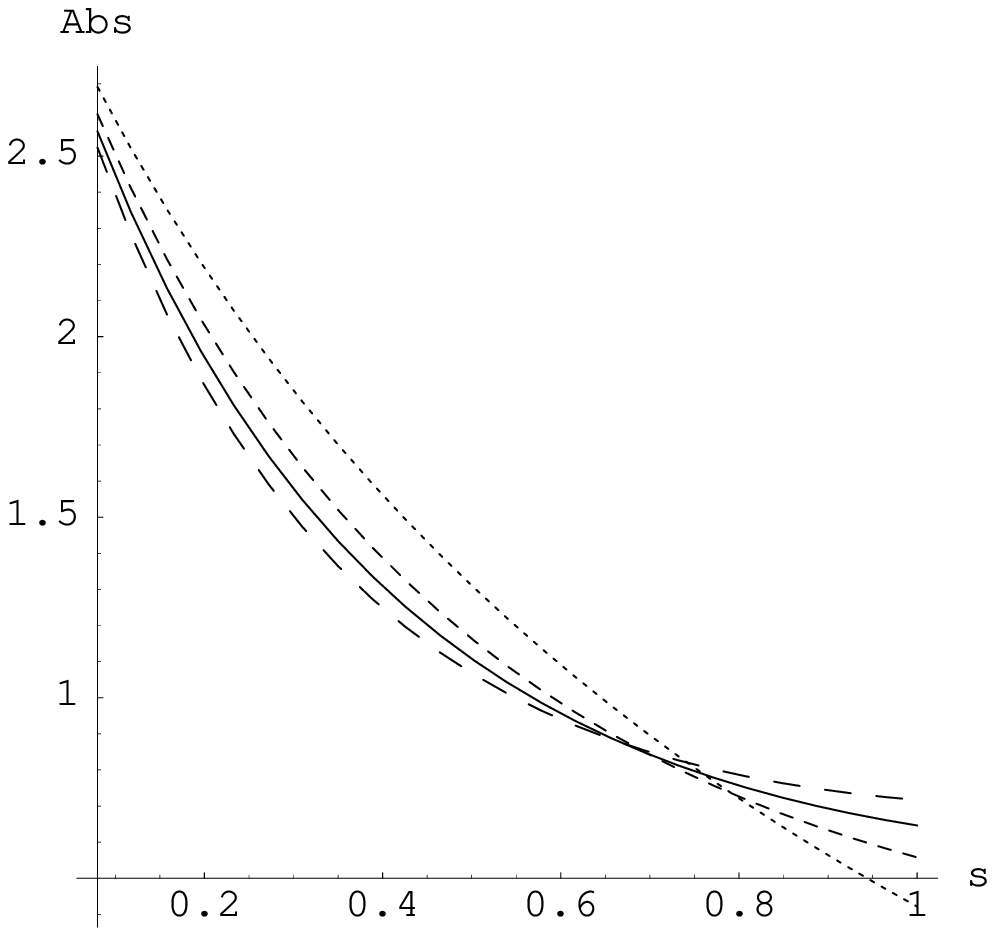}
\caption{}
\label{Fig.6}
\end{figure}

\begin{figure}
\def\picsize{9cm}
\epsfxsize \picsize
\epsfbox{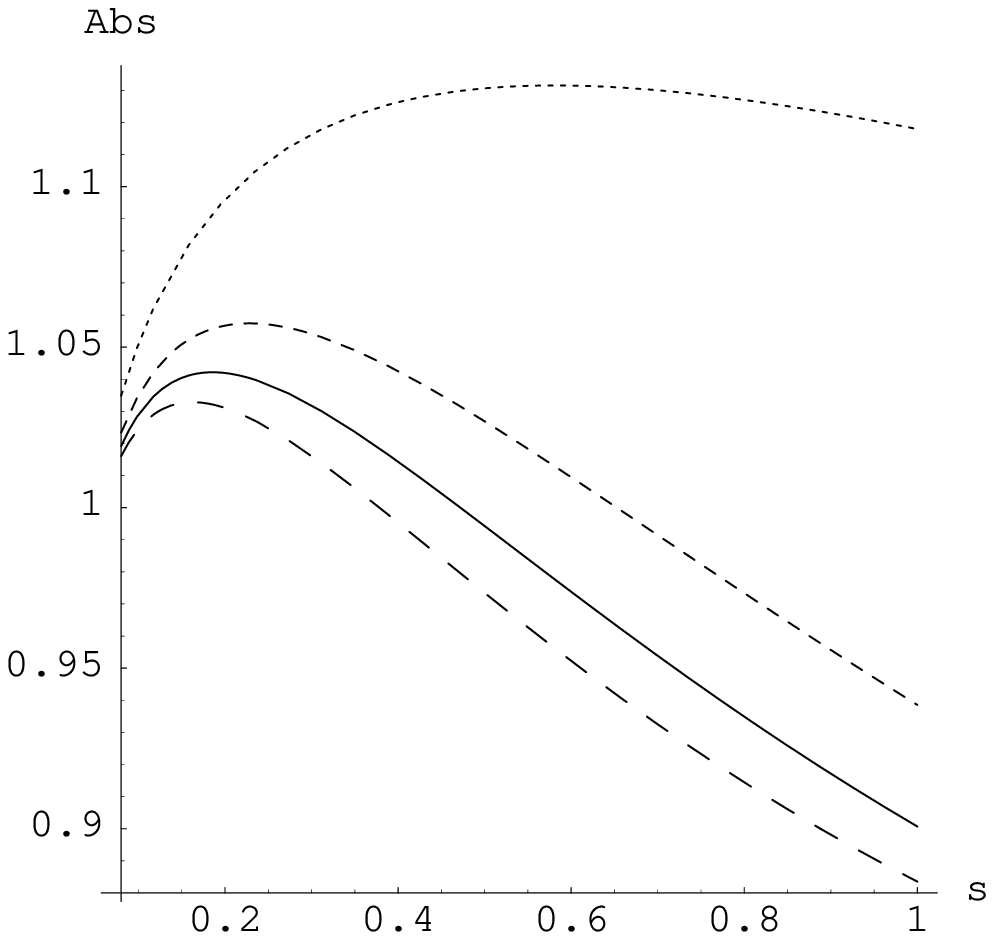}
\caption{}
\label{Fig.7}
\end{figure}

\newpage
\begin{figure}
\def\picsize{9cm}
\epsfxsize \picsize
\epsfbox{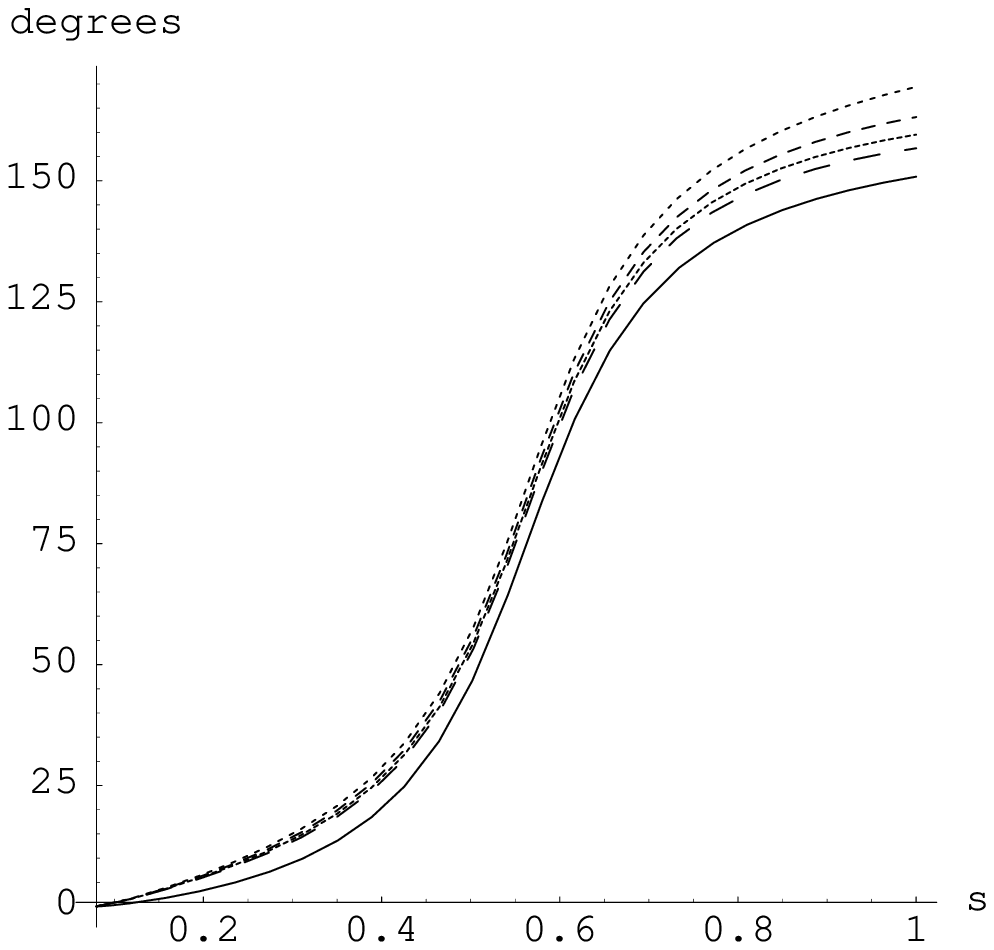}
\caption{}
\label{Fig.8}
\end{figure}

\begin{figure}
\def\picsize{9cm}
\epsfxsize \picsize
\epsfbox{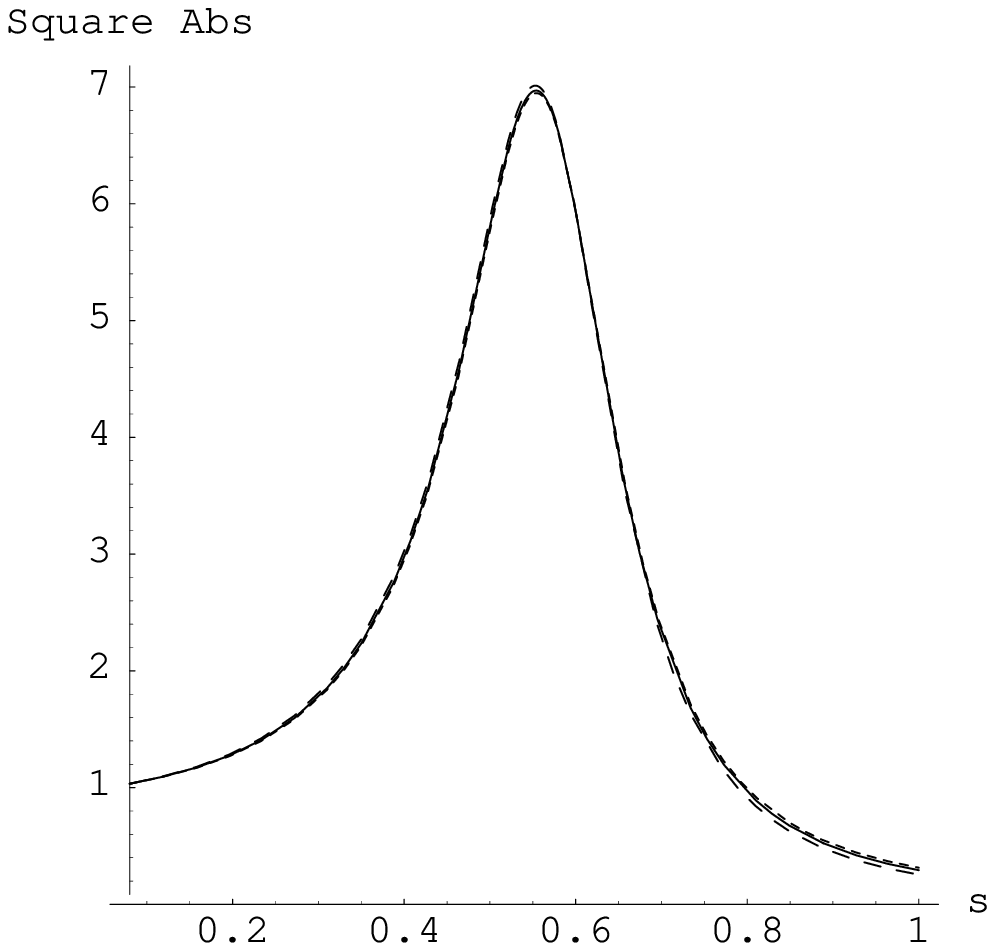}
\caption{}
\label{Fig.9}
\end{figure}

\newpage
\begin{figure}
\def\picsize{9cm}
\epsfxsize \picsize
\epsfbox{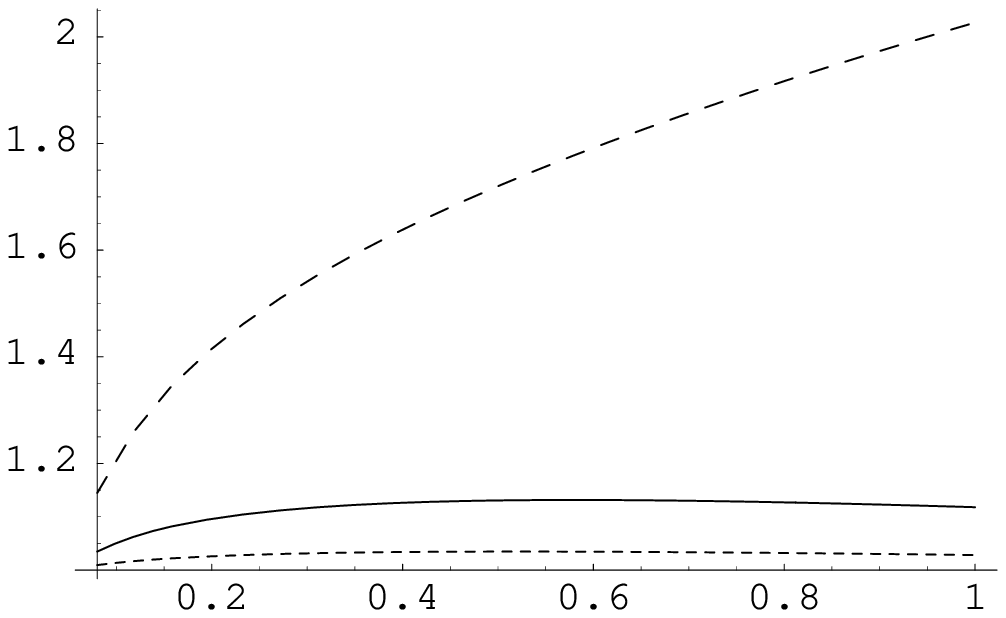}
\caption{}
\label{Fig.10}
\end{figure}


\newpage

\begin{thebibliography}{99}

\bibitem{adler1} S. L. Adler, Phys. Rev.{\bf177}, 2426 (1969); J. S. Bell and R. Jackiw 
Nuovo Cimento {\bf 60A}, 47 (1969).
\bibitem{pdg} Particle Data Group, Eur. Phys. J. C {\bf3}, 1 (1998).
\bibitem{adler2} S. L. Adler, B. W. Lee, S. B. Treiman and A. Zee, Phys. Rev. D {\bf 4}, 
3497 (1971); M. V. Terent'ev, Phys. Lett. {\bf 38B}, 419 (1972); J. Wess and B. Zumino, 
Phys.Lett. {\bf 37B}, 95 (1971); R. Aviv and A. Zee, Phys. Rev. D {\bf 5}, 2372 (1972).
\bibitem{truong1} T. N. Truong, hep-ph/0101345.
\bibitem{antipov} Y. M. Antipov {\it et al.}, Phys. Rev. D {\bf36}, 21 (1987); S. R. 
Amendolia {\it et al.}, Phys. Lett. {\bf 155B}, 457 (1985).
\bibitem{rudaz} S. Rudaz, Phys. Rev. D {\bf 10}, 3857 (1974); Phys. Lett. {\bf 145B},
281 (1984); M. T. Terentev, Phys. Lett. {\bf 38B}, 419 (1972); O. Kaymakcalan, S. Rajeev and J. Schecter, Phys. Rev. D
{\bf 30}, 594 (1984); T. D. Cohen, Phys. Lett. B {\bf233}, 467 (1989).
\bibitem{bando} M. Bando, T. Kugo and K. Yamawaki, Phys. Rep.{\bf
    164}, 217 (1988).T. Fujiwara, T. Kugo, H. Terao, S. Uehara and
  K. Yamawaki, Prog. Theor. Phys. {\bf 73} 926 (1985). 
\bibitem{sharp} M. Gell-Mann, D. Sharp and W. G. Wagner, Phys. Rev. Lett. {\bf 8}, 261 (1962).
\bibitem{bijnens} J. Bijnens, A. Bramon, and F. Cornet, Phys. Lett. B {\bf 237}, 488(1990).
\bibitem{holstein} B. R. Holstein, Phys. Rev. D {\bf53}, 4099 (1996).
\bibitem{hannah} T. Hannah,  Nucl. Phys. {\bf B593}, 577 (2001).
\bibitem{omnes} N. I. Muskhelishvili, {\it Singular Integral Equations}, (Noordhoff, Groningen,
1953);  R. Omn{\`e}s, Nuovo Cimento {\bf 8}, 316 (1958)..

\bibitem{gourdin} M. Gourdin and A. Martin, Nuovo Cimento {\bf16}, 78 (1960); H. S. Wong, Phys. Rev.
Lett. {\bf 5}, 70 (1960).
\bibitem{sakurai} J. J. Sakurai, Ann. Phys. (N.Y) {\bf11}, 1 (1960); M. Gell-Mann and F.
Zachariasen, Phys. Rev. {\bf124}, 953 (1961); Y. Nambu and J. J. Sakurai, Phys. Rev. Lett.
{\bf8}, 79 (1962); {\bf8}, 191(E) (1962).
\bibitem{wagner} M. Gell-Mann, D. Sharp and W. G. Wagner, Phys. Rev. Lett. {\bf 8} 261 (1962).
\bibitem{gounaris} G. J. Gounaris and J. J. Sakurai, Phys. Rev. Lett. {\bf 21}
 24 (1968).
\bibitem{KSRF} K. Kawarabayashi and M. Suzuki, Phys. Rev. Lett. {\bf 16}, 255 (1966);
 Riazuddin and Fayyazuddin, Phys. Rev. {\bf 147}, 1071 (1966).
\bibitem{truong2} T. N. Truong, Phys. Rev. Lett. {\bf61}, 2526 (1988).
\bibitem{watson}  K. M.
Watson, Phys. Rev. {\bf 95}, 228 (1955).cd
\bibitem{truong4}  Le viet Dung and T. N. Truong, Report No hep-ph/9607378.
\bibitem{truong3} T. N. Truong,  hep-ph/0102300.
\bibitem{roy} S. M. Roy, Phys. Lett. {\bf36} B, 353 (1971), J. L. Basdevant, J. C. Le Guillou and H.
Navelet, Nuovo Cimento {\bf7} A, 363 (1972).
\bibitem{brown} L. S. Brown and R. L. Goble, Phys. Rev. Lett. {\bf 20}, 346 (1968).
\bibitem{capraro} L. Capraro \emph{et. al}., Nucl. Phys. {\bf B288}, 659 (1987).
\bibitem{huston} J. Huston
\emph{et. al}., Phys. Rev. D {\bf 33}, 3199 (1986).
\bibitem{bernstein} A. M. Bernstein, Data quoted by Bernstein in his talk given at the Chiral Dynamics meeting
2000., JLAB E94-015
\bibitem{dolinsky} S. I. Dolinsky, \emph{et. al}., Z. Phys. C {\bf 42} 511 (1989).
\bibitem{proto} S. D. Protopopescu \emph{et. al}., Phys. Rev. D {\bf 7}, 1279 (1973).
\bibitem{hyams} B. Hyams  \emph{et. al}., Nucl. Phys. {\bf B64}, 134 (1973).
\bibitem{martin} P. Eastabrooks and A. D. Martin, Nucl. Phys. {\bf B79}, 301 (1974).
\bibitem{barkov} L. M. Barkov, \emph{et. al}., Nucl. Phys. {\bf B256}, 365 (1985)
\bibitem{aleph} ALEPH Collaboration, R. Barate \emph{et. al}., Z. Phys. C {\bf 76}, 15 (1997).

\end{thebibliography}
\end{document}